\documentclass[12pt]{article}

\usepackage{epsfig}

\usepackage{amssymb}
\def\be{\begin{equation}}
\def\ee{\end{equation}}
\def\ba{\begin{array}{c}}
\def\ea{\end{array}}
\def\p{\partial}

\newcommand{\bea}{\begin{eqnarray}}
\newcommand{\eea}{\end{eqnarray}}

\newcommand{\bbr}{\br\!\br}
\newcommand{\kkt}{\kt\!\kt}

\newcommand{\kt}{\rangle}
\newcommand{\br}{\langle}

\begin{document}

\begin{center}

{\Large \bf

 The problem of coexistence of several
 non-Hermitian observables
 in ${\cal PT}-$symmetric quantum mechanics

  }

\vspace{11mm}

{\bf Miloslav Znojil}, {\bf Iveta Semor\'{a}dov\'{a}}, {\bf
Franti\v{s}ek R{u}\v{z}i\v{c}ka},

\vspace{3mm}

Nuclear Physics Institute CAS, 250 68 \v{R}e\v{z}, Czech Republic,

emails:
 {znojil@ujf.cas.cz}, {semoradova@ujf.cas.cz}, {ruzicka@ujf.cas.cz},

%URL: {http://gemma.ujf.cas.cz/$\sim${}znojil/}

\vspace{5mm}

%and
%
%\vspace{3mm}

{\bf Hafida Moulla} and
{\bf Ilhem Leghrib}

\vspace{3mm}

Laboratoire de Physique Mathematique et Subatomique,
Mentouri University, Constantine, Algeria,

e-mails:
  moullahafida@yahoo.fr and leghribilhem@gmail.com

\vspace{5mm}

\end{center}

\vspace{3mm}

\section*{Abstract}

During the recent developments of quantum theory it has been
clarified that the observable quantities (like energy or position)
may be represented by operators $\Lambda$ (with real spectra) which
are manifestly non-Hermitian in a preselected ``friendly'' Hilbert
space ${\cal H}^{(F)}$. The consistency of these models is known to
require an upgrade of the inner product, i.e., mathematically
speaking, a transition ${\cal H}^{(F)} \to {\cal H}^{(S)}$ to
another, ``standard'' Hilbert space. We prove that whenever we are
given more than one candidate for an observable (i.e., say, two
operators $\Lambda_0$ and $\Lambda_1$) in advance, such an upgrade
{\em need not\,} exist in general.

%
%\textbf{PACS:}
%%24.85.+p, 13.85.Hd, 13.87.-a
%\end{abstract}

\newpage

\section{Introduction}

%\subsection{${\cal PT}$ symmetric, non-Hermitian version of
%Schr\"{o}dinger picture}

The traditional presentation of quantum mechanics (called, by
Messiah \cite{Messiah}, ``Schr\"{o}dinger picture'') has recently
been complemented by the formulation which will be called, for our
present purposes, ``${\cal PT}-$symmetric quantum mechanics''
(PTSQM). The distinctive feature of this innovative formulation of
the theory (which has been made popular by Bender with coauthors
\cite{Carl} and in which ${\cal P}$ stands for parity and ${\cal T}$
for time-reversal) lies in its use of certain manifestly
non-Hermitian operators of observables with real spectra (cf. also
the detailed summaries of the idea in \cite{ali} or in several
slightly more mathematics-oriented concise reports collected in the
newest book \cite{book} on the subject).

One of the most interesting questions connected with the
applicability of the PTSQM formalism was already discussed in the
earlier review by Scholtz, Geyer and Hahne \cite{Geyer}. This
question concerns the acceptability and mutual compatibility of the
{\em two or more} preselected independent non-Hermitian operators of
observables with real spectra. In this perspective our present
contribution to the further development of the formalism found its
immediate predecessor in paper \cite{dvojice}. One of us proposed
there a prototype quantum toy model in which the mathematical
compatibility between the {\em two} non-Hermitian but ${\cal PT}$
symmetric observables (viz., between a non-Hermitian Hamiltonian and
a non-Hermitian spin projection) has been guaranteed by their mutual
commutativity.

Over-restrictive as the commutativity constraint certainly is, it
found its inspiration in a few other commutativity-requiring
constructions where the pairs of non-Hermitian observables were
specified as the Hamiltonian plus the so called quasi-parity
\cite{2001} or as the Hamiltonian plus the so called charge
\cite{BBpra,mz2}. Naturally, the situation in which one would have
to deal with some two (or more) {\em entirely independent}
non-Hermitian candidates for observables (say, $\Lambda_1$ and
$\Lambda_2$, with real spectra) would be much more interesting.

In our present paper we intend to return to the problem and to
reanalyze the operator-pair compatibility problem in the form in
which the commutativity constraint $[\Lambda_1,\Lambda_2]=0$ is
replaced by a more sophisticated {\it ad hoc} condition (cf., e.g.,
the samples of such non-commuting pairs of observables in the series
of Refs.~\cite{jana,protijane,BF,Fring}).

We believe that up to now, a more systematic, model-independent
study of the non-Hermitian compatibility problem was not published.
This gap is to be filled in what follows. In particular, we shall
demonstrate that there exists a subtle correspondence between the
choice of the so called irreducible sets of the candidates
$\Lambda_j$ for the observables and the role played by these sets in
the removal of the well known ambiguity of the assignment of the
Schr\"{o}dinger-picture Hilbert space ${\cal H}^{(S)}$ to a {\em
single} observable $\Lambda_0$ (cf., e.g., Ref.~\cite{SIGMAdva} for
a rather formal but fairly exhaustive technical discussion of the
latter topic).

The presentation of our results will be preceded by a concise
outline of the PTSQM formalism in section \ref{theodor}. The
formulation of the non-Hermitian compatibility problem will follow
in section \ref{se3}. For illustration we mention there the
triviality of the case in which the two preselected observables $H$
(or, more generally, any non-Hermitian operator $A$ {\it alias}
$\Lambda_0$ with real spectrum) and $X$ ({\it alias} $B$ or
$\Lambda_1$) commute (subsection \ref{se3a}). Subsequently, in
subsection \ref{secabove} we deliver the proof of the
incompatibility of the entirely generic $\Lambda_0$ with the
entirely generic $\Lambda_1$ (which does not commute with the former
one of course), assuming only, for simplicity, that the Hilbert
spaces in question are finite-dimensional.

Basically, the latter result means that the well known assignment of
a proper probabilistic interpretation to a {\em single} preselected
non-self-adjoint $N$ by $N$ matrix of an observable $A\neq
A^\dagger$ need not necessarily admit the acceptability of another,
independent $N$ by $N$ matrix of another candidate $B\neq B^\dagger$
for an observable in the same quantum system. We will be able to
conclude that in a generic situation, the arbitrarily selected
doublet of non-Hermitian operators $A$ and $B$ with real spectra
{\em cannot} be immediately treated as representing a pair of the
observable quantities. For a coexistence, the operators must be
nontrivially interconnected.

The discussion of the practical applicability aspects of our present
contribution is initiated in section \ref{se4}. We recall there the
weakly $q-$deformed version of one of the above-mentioned models
(subsection \ref{sec0}). Next we show how the required compatibility
conditions become simplified in the first-order perturbation
approximation (subsection \ref{sec0be}). Subsequently (i.e., in
section \ref{tri}) a few technical difficulties associated with the
construction of the metric itself are discussed for the single
generic input observable $\Lambda_0$ (subsection \ref{sec1a}) and in
the situation in which one adds another operator, $\Lambda_1$
(subsection \ref{sec1b}). Section \ref{pet} is finally devoted to a
few mathematical subtleties, i.e., more explicitly, to the role of
the boundedness of the operators of observables (cf. subsection
\ref{petar}), to the merits of the combination of the perturbation
and truncation strategies (listed in subsection \ref{petas}) and to
the explicit and exhaustive illustrative description of the
compatibility criteria at $N=2$ in subsection \ref{petant}. The
summary of our message is finally provided by section \ref{summary}.

\section{A concise outline of the theory\label{theodor}}

\subsection{The Dyson's quantum mechanics {\it in nuce}}

The PTSQM formalism should be perceived as fully compatible with the
traditional textbooks on quantum mechanics \cite{Messiah}. A more
detailed explanation of this point of view may be found summarized
in papers \cite{timedep} or reviews \cite{ali,Geyer,SIGMA}. In a way
inspired also by Refs.~\cite{Dyson,Dieudonne} we will characterize
the PTSQM as a representation of quantum systems in which the use of
the ``standard'' Hilbert space of states ${\cal H}^{(S)}$ (in which
the observables are represented by the self-adjoint operators,
$\Lambda = \Lambda^\ddagger$) can be {\em paralleled} by the use of
another, unphysical but mathematically friendlier Hilbert space
${\cal H}^{(F)}$. In the latter space the {\em same} observables
appear non-Hermitian of course, $\Lambda \neq \Lambda^\dagger$, due
to the underlying auxiliary {\em intentional simplification} of the
inner product.

\begin{table}[h]
\caption{Non-Hermitian version of Schr\"{o}dinger picture (cf.
\cite{SIGMA}). } \label{unota}

\vspace{2mm}

\centering
\begin{tabular}{ccccc}
\hline \hline \multicolumn{5}{c}{the
triplet of representation Hilbert spaces:}\\
   $ {\cal H}^{(F)}$ && $ {\cal H}^{(S)}$
    && $ {\cal H}^{(T)}$ \\
 = the first one,&&= the second one,&&= the third one,\\
 false space,&& standard space &&of textbooks,\\
 unphysical&& &&inaccessible\\
 \hline
 \hline \multicolumn{5}{c}{the purpose of their simultaneous use:}\\
  {the friendlier}  &  &\fbox{the correct space} &
 &  {the friendlier}  \\
   mathematical  & $\swarrow $ &&$ \searrow$
 &  probabilistic  \\
   \fbox{representation}  &  &&
 & \fbox{interpretation} \\
        \hline
 \hline \multicolumn{5}{c}{formally, the generic observable
 is, respectively,}\\
   non-Hermitian  &  &self-adjoint&
 &   self-adjoint \\
   ($\Lambda \neq
\Lambda^\dagger$)  &  &($\Lambda = \Lambda^\ddagger$)&
 &   (transformed) \\
 \hline
 \hline
\end{tabular}
\end{table}

The structure of the resulting ``three-Hilbert-space'' (THS)
Schr\"{o}dinger-picture pattern is summarized in Table \ref{unota}.
The introduction of such a representation of a quantum system dates
back to Dyson \cite{Dyson} who conjectured that one need not
distinguish between the predictions made within the standard
physical Hilbert space $ {\cal H}^{(S)}$ and within any other,
unitarily equivalent alternative Hilbert space $ {\cal H}^{(T)}$
``of textbooks''.

In the context of applications the Dyson's recipe was successful in
nuclear physics where one always knows the self-adjoint Hamiltonian
in $ {\cal H}^{(T)}$. Once this operator is found overcomplicated,
one is well motivated to change the Hilbert space as well as to
simplify the Hamiltonian. This may be achieved via certain {\it ad
hoc}, intuition-based and invertible ${\cal H}^{(F)} \to {\cal
H}^{(T)}$ mapping, to be called the Dyson's map and to be denoted by
the dedicated symbol $\Omega$ in what follows.

\subsection{The ${\cal PT}-$symmetric quantum mechanics {\it in nuce}}

%
%\subsection{The single-non-Hermitian-observable problem}

In the  PTSQM setting the old Dyson's recipe was inverted (cf.,
e.g., review paper \cite{Carl} for numerous illustrations). In the
first step one just picks up an auxiliary Hilbert space $ {\cal
H}^{(F)}$ together with a sufficiently elementary (quite often, just
ordinary-differential-operator) non-Hermitian but ${\cal
PT}-$symmetric candidate $H$ for the Hamiltonian.

The preselected Hilbert space ${\cal H}^{(F)}$ is then declared
``false''. For the correct physical interpretation purposes it is to
be replaced by another, ``second'' Hilbert space ${\cal H}^{(S)}$.
In the latter, non-equivalent, $S-$superscripted Hilbert space the
initial non-Hermitian operators $H$ of Hamiltonians must be
self-adjoint. Thus, one might write
 \be
 {H_{}}= {H_{}}^\ddagger\,.
 \label{formu}
 \ee
The acceptance of such a convention could be misleading.
Fortunately, its immediate use is also not necessary because via a
mere redefinition of the inner product we may {\em always} postpone
the study of Eq.~(\ref{formu}) and stay working inside ${\cal
H}^{(F)}$.

The Hermiticity property (\ref{formu}) does not get lost in  ${\cal
H}^{(F)}$. It remains mediated by the replacement of
formula~(\ref{formu}) valid in ${\cal H}^{(S)}$ by {\em the same}
formula re-written (i.e., represented) in ${\cal H}^{(F)}$,
 \be
 {H_{}}= \Theta^{-1}H^\dagger\Theta\,,
 \ \ \ \ \ \ \ \Theta=\Omega^\dagger\Omega\,.
 \label{reformu}
 \ee
The operator $\Theta$ is called the physical Hilbert-space metric.

The preference of Eq.~(\ref{reformu}) suppresses many
misunderstandings. First of all,  we may keep writing the ``false'',
$F-$superscripted inner products in the standard Dirac's bra-ket
notation,
 \be
 [\br \psi_1|\psi_2\kt]^{(F)}\ \equiv \ \br \psi_1|\psi_2\kt\,.
 \ee
Secondly, the representation of the other, physical,
$S-$superscripted inner products may be given virtually equally
friendly form in its friendlier $F-$space representation,
 \be
 [\br \psi_1|\psi_2\kt]^{(S)}\
 \equiv \ \br \psi_1|\Theta|\psi_2\kt\,.
 \ee
In the spirit of review \cite{SIGMA} one can also introduce the
doubled bra and the doubled ket symbols and abbreviate
$\Theta|\psi\kt\,\equiv\,|\psi\kkt$ and $\br
\psi|\Theta\,\equiv\,\bbr\psi|$. Although such a convention is
fairly unusual (and it will not be used too much in the bulk text of
this paper), its use can make the formalism more transparent because
the set of special ketkets $|n_0\kkt$ can be introduced as a set of
eigenvectors of the conjugate Hamiltonian, $H^\dagger|n_0\kkt =
E_n\,|n_0\kkt$. As a consequence one can endow the Hamiltonian
operator $H$ with the menu of {\em all\ } of its eligible
Hermitizing metrics,
 \be
 \Theta=\sum_n\,|n_0\kkt \kappa_n\bbr n_0|=\Theta(\vec{\kappa})\,.
 \label{muta}
 \ee
The coefficients $\kappa_n$ are variable and the sequence of their
values must be kept real, positive and properly bounded
\cite{SIGMAdva}.

\section{The operator-compatibility conditions\label{se3}}

\subsection{The case of commuting pairs of non-Hermitian observables \label{se3a}}

The Hilbert-space-metric operator $\Theta$ (with multiple necessary
mathematical properties \cite{Geyer,Siegl,sBag,aliSkot,ATbook} which
were not mentioned here for the sake of brevity) enables one to pull
back the $S-$space formula (\ref{formu}) to its explicit equivalent
representation (\ref{reformu}) in $F-$space. Naturally, the
correspondence between the left eigenvectors and the general
Hermitizing metric operators applies to the Hamiltonian
$H=\Lambda_0$ as well as to any other non-Hermitian candidate
$\Lambda_j$ for the operator of an observable. Via an analogue of
formula (\ref{muta}), any such an operator can be assigned, {\it
mutatis mutandis}, an exhaustive menu of eligible metrics
 \be
 \Theta=\sum_n\,|n_j\kkt \kappa_n^{(j)}\bbr n_j|
 =\Theta_j(\overrightarrow{\kappa^{(j)}})\,.
 \label{remuta}
 \ee
Our present research subject can briefly be characterized, in its
first nontrivial version with $j=0$ and $j=1$ in Eq.~(\ref{remuta}),
as the analysis of the operator twin-observability constraint
 \be
 \Theta_0(\overrightarrow{\kappa^{(0)}})=
 \Theta_1(\overrightarrow{\kappa^{(1)}})\,.
 \label{bota}
  \ee
Just the trivial versions of the solution of this relation are
usually considered in the literature. Once we recall that the shared
eigenvalues $E_n$ of $\Lambda_0=H$ and of
$\Lambda_0^\dagger=H^\dagger$ are, by assumption, real, we may
decide to complement our knowledge of the eigenketkets $|n_0\kkt$ of
the conjugate $H^\dagger$ by the standard eigenkets $|n_0\kt$ of
$H$. In this way we obtain a biorthonormalized basis \cite{biorto}
and a spectral-like formula
 \be
 H = \sum_n\,|n_0\kt\,E_n\,\bbr n_0|\,.
 \label{deweaker}
 \ee
If we introduce another operator
 $\Lambda_1\neq
\Lambda_1^\dagger$ with real spectrum and such that
 \be
 \Lambda_1 = \sum_n\,|n_1\kt\,b_n\,\bbr n_1|\,,
 \label{reweaker}
 \ee
the related biorthonormal basis will be different because the new
operator does not commute with the old one in general.

Once we now impose the simplifying condition of the commutativity,
i.e., of the coincidence of the two eigenbases in formulae
 \be
 \Lambda_0^{(special)} = \sum_n\,|n\kt\,a_n\,\bbr n|\,,
 \ \ \ \ \ \
 \Lambda_1^{(special)} = \sum_n\,|n\kt\,b_n\,\bbr n|\,
 \label{weaker}
 \ee
we find that such an assumption immediately converts the
compatibility condition~(\ref{bota}) into an identity. In the light
of Appendix of Ref.~\cite{Geyer} the introduction of the new
observable does not remove the ambiguity from the metric
(\ref{muta}) at all. The free variability of the whole multiplet of
parameters $\vec{\kappa}$ survives. In the terminology of review
\cite{Geyer}, the set of the two observables (\ref{weaker}) remains
reducible.

%
%\subsection{The conditions of coexistence of the two generic non-Hermitian
%observables defined in a finite, $N-$dimensional Hilbert space
%${\cal H}^{(F)}$\label{ctyri}}

% for a given pair $\Lambda_0$ and
%$\Lambda_1)$  of the input operators (\ref{deweaker}) and
%(\ref{reweaker}). Marginally
% \be
% \Lambda_0 \to
%\Theta^{}_0(\overrightarrow{\kappa^{(0)}})=
%\Theta^{(A)}(\overrightarrow{\alpha})\,,\ \ \ \ B \to
%\Theta^{(B)}(\overrightarrow{\kappa^{(1)}})=
%\Theta^{(B)}(\overrightarrow{\beta})
% \ee

%\subsection{Non-commuting non-Hermitian observables }

Let us now exclude the fully degenerate scenario of
Eq.~(\ref{weaker}) as trivial and let us assume that the two
biorthonormal bases entering spectral expansions (\ref{deweaker})
and (\ref{reweaker}) do not coincide. The operator-coincidence
(\ref{bota}) will then restrict the variability of the coefficients
in Eq.~(\ref{muta}). The number of the constraints which are
generated by Eq.~(\ref{bota}) becomes {\em too large} for the
purpose in general.

Certainly, the problem deserves a deeper, more concrete critical
analysis.

\subsection{The case of non-commutative non-Hermitian
observables \label{secabove}}

{\it A priori} we may expect that the metric
$\Theta(\Lambda_0,\Lambda_1)$ can only exist under certain
constraints represented, formally, by the solvability of
Eq.~(\ref{bota}). The knowledge of these constraints will be
difficult to extract in practice. This makes our present task (i.e.,
the search for criteria) rather nontrivial.

During the search for a guarantee of the existence of {at least one
acceptable metric} we reveal that in contrast to the comparatively
popular suppressions of the ambiguity of $\Theta$, the danger of the
nonexistence of {\em any} metric is often underestimated in the
literature. For example, the authors of Ref.~\cite{BF} took the
existence of $\Theta(\Lambda_0,\Lambda_1)$, rather naively, for
granted. In an entirely general setting they considered ``a definite
form of the Hamiltonian $H$'' (i.e., in our notation, of the first
operator $\Lambda_0$) ``and an additional observable'' (i.e.,
another operator $\Lambda_1$). They claimed, {\it expressis verbis},
that ``these two choices \ldots will fix the metric uniquely, such
that there are no ambiguities left in the interpretation of the
physical observables''~\cite{BF}.

For a technically feasible critical analysis of such a claim let us,
first of all, skip certain less essential mathematical subtleties
and let us restrict attention to the models living in
finite-dimensional Hilbert spaces, specifying, for the sake of
definiteness, ${\cal H}^{(F)}=\mathbb{C}^N$. In this case we may use
{\em any} suitable basis and we may treat our operators $\Lambda_0$,
$\Lambda_1$, \ldots as the respective complex $N$ by $N$ matrices
$A=A^{(N)}$, $B=B^{(N)}$, \ldots .

Once we abbreviate, in parallel,
$\overrightarrow{\kappa^{(0)}}=\overrightarrow{\alpha}$ and
$\overrightarrow{\kappa^{(10)}}=\overrightarrow{\beta}$, the
respective metrics in Eq.~(\ref{bota}) may be perceived as the well
defined Hermitian-matrix functions of the real and positive
variables. This enables us to rewrite Eq.~(\ref{bota}) as the
algebraic set
 \be
 \Theta_{mn}^{(A)}(\overrightarrow{\alpha})=
 \Theta_{mn}^{(B)}(\overrightarrow{\beta})
 \,,
 \ \ \ \ \
 m,n=1,2,\ldots,N
 \,
 \label{robota}
  \ee
of $N^2$ linear equations. The left-hand-side matrix with
eigenvalues $\theta^{(A)}_n(\overrightarrow{\alpha})$ may be
diagonalized by a suitable unitary matrix
$U(\overrightarrow{\alpha})$. This yields an equivalent set of
relations
 \be
 \theta^{(A)}_n(\overrightarrow{\alpha})\delta_{mn}=
 \left [U(\overrightarrow{\alpha})
 \,\Theta^{(B)}(\overrightarrow{\beta})
 U^\dagger(\overrightarrow{\alpha})
 \right ]_{mn}\,,
 \ \ \ \ m,n=1,2,\ldots, N\,.
 \label{zarobotok}
  \ee
For an arbitrary ``input'' choice of the $N-$plet of variable
parameters $\alpha_n$ this certainly provides the $N-$plet of
constraints
 \be
 \theta^{(A)}_n(\overrightarrow{\alpha})=
 \left [U(\overrightarrow{\alpha})
 \,\Theta^{(B)}(\overrightarrow{\beta})
 U^\dagger(\overrightarrow{\alpha})
 \right ]_{nn}\,,
 \ \ \ \ n=1,2,\ldots, N\,
 \label{kozarobotok}
  \ee
which may be read as an implicit-function definition of the
``output'' $N-$plet of quantities
$\beta_n=\beta_n(\overrightarrow{\alpha})$.

We are now left with the remaining independent $N(N-1)/2$ conditions
forming the upper triangular matrix in (\ref{zarobotok}) and
reflecting the presence of the left-hand-side zeros,
 \be
  \left [U(\overrightarrow{\alpha})\,
  \Theta^{(B)}(\overrightarrow{\beta}(\overrightarrow{\alpha}))
  U^\dagger(\overrightarrow{\alpha})
 \right ]_{mn}=0\,,
 \ \ \ \ m=n+1,n+2,\ldots, N\,,
 \ \ \ n = 1, 2, \ldots, N-1\,.
 \,.
 \label{nezarobotok}
  \ee
Our remaining available $N-$plet of the real and positive variables
$\overrightarrow{\alpha}$ is constrained by the overdetermined set
of the $N(N-1)/2$ complex (i.e., of the $N(N-1)$ real) nonlinear
algebraic equations. In the generic case one can use the roughmost
estimate of the number of solutions and conclude that at $N> 2$ the
nontrivial real and positive roots $\alpha_n$ need not exist at all.
Indeed, the number of equations (\ref{nezarobotok}) will not exceed
the number of unknowns only when $N(N-1)\leq N$, i.e., at $N \leq
2$.

In the opposite direction the requirement of the existence of at
least one real and positive $N-$plet of parameters $\alpha_n$
necessarily imposes certain nontrivial restrictions upon our freedom
of the choice of the ``dynamical input'' operators $A$ and $B$.
Naturally, the latter set of restrictions becomes perceivably more
stringent when one decides to accommodate more than two ``dynamical
input'' operators $\Lambda_j$ with $j\leq j_{max}>1$. At the same
time, our theoretical ambitions should not be exaggerated. Thus, for
$j_{max}\gg 1$ at least, it seems to make good sense to follow the
common practice of the traditional textbooks where, typically, a
trivial unit-matrix metric $\Theta=I$ is chosen and fixed in
advance.

In the present generalized setting the acceptance of the same
philosophy will merely mean that one simplifies the strategy and
picks up and fixes a suitable nontrivial initial matrix
$\Theta^{(A)}\neq I$. In this framework one then admits only those
additional observables which obey the Dieudonn\'{e}'s
quasi-Hermiticity condition
 \be
 \Lambda_j\Theta^{(A)} =\Theta^{(A)}\,\Lambda_j^\dagger
 \ee
in ${\cal H}^{(F)}$ (cf.~Eq.~(\ref{reformu})), i.e., the
hidden-Hermiticity requirement $\Lambda_j=\Lambda_j^\ddagger$ in the
fixed physical Hilbert space ${\cal H}^{(S)}$.

\section{Applications of the theory\label{se4}}

The {\em sufficient} relations as sampled by Eq.~(\ref{weaker})
might prove insufficiently general. The less restrictive and more
general $N \leq \infty$ formula (\ref{nezarobotok}) is too implicit
for being useful in practice. Only the study of specific toy models
can lead to some more definite conclusions.

\subsection{Working with weakly deformed oscillator algebras\label{sec0}}

The uncertainty relations for positions and momenta acquire rather
unusual forms after a tentative replacement of the solvable harmonic
oscillator by its suitable $q-$deformed analogues \cite{origins}. In
a way related to the Connes' phenomenology-oriented conjecture of
making the geometry non-commutative \cite{1}, these developments had
a perceivable impact also upon quantum physics \cite{BF,2}.

A typical build up of quantum models of this type starts from
Heisenberg algebra. A coordinate $x$ and momentum $p\ $ or,
alternatively, an annihilation operator $a$ and a creation operator
$a^+$ may be deformed, typically, in such a way that $(a,a^+ ) \to
(\widehat{a}, \widehat{a}^+)$ and
 \begin{equation}
 \widehat{a}\widehat{a}^{+}-q\widehat{a}^{+}%
 \widehat{a}
 \ \equiv \ \left[ \widehat{a},\widehat{a}^{+}\right] _{q}=1\,.
 \label{lola}
 \end{equation}
Some of the complications become simplified in a vicinity of the
zero-deformation limit $q=1$, i.e., at small
$\varepsilon=q-1\neq 0$. One may recall Eq.~(\ref{lola}), ignore the
higher-order corrections and construct, in a simplified
leading-order approximation,
 \begin{equation}
\widehat{a}=\frac{1}{\sqrt{2}}\left(
\widehat{X}_{q}+i\widehat{P}_{q}\right) =a+\frac{1
}{4}\,\varepsilon\,a^{+}a^{2} + {\cal O}(\varepsilon^2)\,,
\end{equation}
 \begin{equation}
 \widehat{a}^{+}=\frac{1}{\sqrt{2}}
 \left( \widehat{X}_{q}-i\widehat{P}_{q}\right)
 =a^{+}+\frac{1 }{4}\,\varepsilon\,a^{+^{2}}a + {\cal
 O}(\varepsilon^2)\,.
 \end{equation}
One may then rewrite the Hamiltonian of the deformed harmonic
oscillator in its perturbation-approximation form
 \begin{equation}
 A=\widehat{H_{q}}=\widehat{P}_{q}^{2}+\widehat{X}_{q}^{2}
 ={H}_{0}+\frac{1 }{8}\,\varepsilon\,{H}_{1}
 + {\cal O}(\varepsilon^2)\,.
 \label{a1}
\end{equation}
The standard textbook harmonic oscillator Hamiltonian
$H_{0}=p^{2}+x^{2}=H_0^\dagger $ is, in the leading-order
approximation, complemented by a manifestly non-selfadjoint {\it
alias} non-Hermitian interaction operator
\begin{equation}
H_{1}=2x^{4}-x^{2}+3p^{2}-3+2ix^{3}p+2ixp^{3}+2x^{2}p^{2}-8ixp\,
\neq \,H_1^\dagger
  \,. \label{a5}
 \end{equation}
Such a model was studied in \cite{post1} and its spectrum was found,
in the leading-order approximation, real. This opened the question
of the possible coexistence of the observability, say, of the
Hamiltonian (i.e., operator $A=\widehat{H_{q}}$) and of the
coordinate (operator $B=\widehat{X_{q}}$ -- cf. also Refs.~\cite{BF}
and \cite{post2} in a slightly different context).

\subsection{The perturbative Hermitizability of $A \neq
A^\dagger$ and  $B \neq B^\dagger$\label{sec0be}}

The study of the $q-$deformed Hamiltonian $A=\widehat{H_{q}} \neq
A^\dagger$ originated from its interesting spatial {geometry}
features \cite{2,BF2}. The Hermitization of the Hamiltonian {\em
must} be accompanied by the equally relevant Hermitization of the
operator of the coordinate
 \be
 B=\widehat{X}_{q}=x+\frac{1 }{8}\,\varepsilon\,\left(
 x^{3}-xp^{2}+ix^{2}+ix^{2}p-x+ip^{3}+pxp+p^{2}x\right) + {\cal
 O}(\varepsilon^2)\,.
 \label{a}
 \ee
We may even try to work with a multiplet of observables
$\widehat{\Lambda}_j$ where $\widehat{\Lambda}_0=A=H$,
$\widehat{\Lambda}_1=B$, etc. All of these operators of observables
must satisfy the observability constraint,
 \be
 \widehat{\Lambda}_j= \widehat{\Lambda}^\ddagger_j
  \ \equiv \ \Theta^{-1}  \widehat{\Lambda}^\dagger_j\
 \Theta\,,
 \ \ \ \ \
 j= 0, 1,\ldots, j_{max}
 \,.
 \label{constra}
 \ee
These relations must {\em all\ }\ contain {\em the same} physical
metric operator $\Theta$.

%\subsection{An approximative, perturbation-approximation
%analysis\label{4.2}}

In the limit of infinitesimally small non-Hermiticities it is
nontrivial to guarantee the existence of at least one formal set of
quantities $\beta_n$, $\alpha_n$, $\Theta$, $A$ and $B$ which would
be compatible with the twin-observability condition~(\ref{robota}).
In the leading-order approximation one can consider, in the spirit
of Eqs.~(\ref{a1}) and (\ref{a}), the $N$ by $N$ matrices
 \be
 A=A_0+\varepsilon\,A_1+\ldots\,,\ \ \ \
 A_0=A_0^\dagger\,
 \label{21}
 \ee
and
 \be
 B=B_0+\varepsilon\,B_1+\ldots\,,\ \ \ \
 B_0=B_0^\dagger\,
 \label{22}
 \ee
of the relevant observables. The latter ansatz could be generalized
to a higher-order precision and/or to a larger number of operators
$j_{max}>1$. For its current $j_{max}=1$ form, our task may be now
formulated as the construction of the leading-order metric
 \be
 \Theta=I + \varepsilon\,F+\ldots\,,
 \ \ \ \ \ \ F=F^\dagger\,.
 \ee
The latter operator must make {\em both} of the input operators $A$
and $B$  of observables compatible, within given precision, with the
respective hidden-Hermiticity constraints (\ref{constra}). Thus, the
following two commutator-containing equations
 \be
 A_0\,F-F\,A_0=A_1-A_1^*\ \equiv\ {\rm i}\,R\,,\ \ \ \ \ \
 R=R^\dagger\,,
 \label{16}
 \ee
 \be
 B_0\,F-F\,B_0=B_1-B_1^*\ \equiv\ {\rm i}\,S\,,\ \ \ \ \ \
 S=S^\dagger\,
 \label{17}
 \ee
are to be solved for the unknown complex first-order metric-operator
component $F=F^\dagger$. Due to the use of the perturbation theory,
their form is perceivably simpler than that of their
non-perturbative predecessors (\ref{bota}) and (\ref{robota}).

\section{The perturbative construction of
the metric\label{tri}}

\subsection{The single-observable problem\label{sec1a}}

Without a pre-selected $\Theta$ the constructive treatment of the
observability constraints (\ref{constra}) is difficult \cite{117}.
Fortunately, people are often interested only in the observability
of the Hamiltonian,
 \be
 \widehat{H_{q}}=  \Theta^{-1}  \widehat{H_{q}}^\dagger
 \Theta\,.
 \label{formulka}
 \ee
The analysis of this equation may be facilitated by additional
assumptions. The restriction to bounded operators is the most
important one. Another postulate, remarkably efficient in
applications, introduces the auxiliary ${\cal PT}-$symmetry property
which is equivalent to the relation $\widehat{H_{q}}^\dagger {\cal
P}={\cal P}\widehat{H_{q}}$ where ${\cal P}$ denotes the operator of
parity.

Once we turn attention from the Hamiltonian of Eq.~(\ref{formulka}),
say, to the coordinate of Eq.~(\ref{a}) with property
\begin{equation}
\widehat{X}_{q}^{+}=\widehat{X_{q}}+\frac{\varepsilon }{8}\left(
2xp^{2}-ix^{2}p-2ip^{3}-2p^{2}x-ipx^{2}\right) + {\cal
O}(\varepsilon^2) \label{a24}
\end{equation}
we may recall constraint (\ref{constra}), i.e.,
\begin{equation}
X_{q}^{+}=\Theta X_{q}\Theta ^{-1}  \label{a25}
\end{equation}
and we may search for such a form of the metric which would be
positive, invertible and factorizable,
\begin{equation}
\Theta =\Omega^{+}  \Omega\,.  \label{a26}
\end{equation}
One of such formal solutions can be found and expressed as an
exponential,
\begin{equation}
\Theta_0 =e^{\varepsilon f\left( x,p\right)+ {\cal O}(\varepsilon^2)
}= 1+\varepsilon f\left( x,p\right)+ {\cal O}(\varepsilon^2)
\label{a27}
\end{equation}
with
\begin{equation}
f\left( x,p\right) =\frac{1}{4}\left[ \frac{1}{4}\left(
x^{2}p^{2}+p^{4}+p^{2}x^{2}\right) +\frac{i}{3}\left(
xp^{3}-p^{3}x\right) \right]\,.  \label{a31}
\end{equation}
Such a solution signals several warnings at once. It is not
acceptable, first of all, because of its unboundedness. In the next
paragraph we will pay more attention to the consistency between
Eqs.~(\ref{a25}) and (\ref{a27}). We shall show that nontrivial as
it is,  this solution cannot render the Hamiltonian selfadjoint.

\subsection{Tentative Hermitizations and their failures\label{sec1b}}

According to the expectations as expressed in
Refs.~\cite{jana,protijane,BF,Fring,post1} the
single-observable construction of preceding paragraph should suffice
for the necessary Hermitization of the Hamiltonian. According to the
same sources the use of the most common special mapping
\begin{equation}
\widehat{H_q} \ \to \ \mathfrak{h}_0=\Omega_0 \widehat{H_q}\Omega_0
^{-1} \label{a30}
\end{equation}
with the widely recommended choice of the square-root form of
$\Omega_0=\sqrt{\Theta_0}$ {\em might} yield a {\em manifestly
Hermitian} partner Hamiltonian $\mathfrak{h}_0 =
\mathfrak{h}_0^\dagger$ acting in ${\cal H}^{(T)}$.

Although the analysis of this hypothesis is far from easy, the
straightforward evaluation of the difference  $\triangle
=\mathfrak{h}_0 - \mathfrak{h}_0^\dagger$ falsifies the hypothesis,
$\triangle \neq 0$. The validity of such a proof was also
reconfirmed via its computer-assisted re-verification \cite{Tater}.
One can conclude that the coordinate-Hermitizing Dyson-mapping
operator
\begin{equation}
\Omega_0 =e^{\varepsilon g\left( x,p\right) + {\cal
O}(\varepsilon^2)}= 1+\varepsilon g\left( x,p\right)+ {\cal
O}(\varepsilon^2) \label{a28}
\end{equation}
with
\begin{equation}
g\left( x,p\right) =\frac{1}{8}\left[ \frac{1}{4}\left(
x^{2}p^{2}+p^{4}+p^{2}x^{2}\right) +\frac{i}{3}\left(
xp^{3}-p^{3}x\right) \right]  \label{a32}
\end{equation}
{\em does not} Hermitize the Hamiltonian. This disproves the
hypothesis and reopens the methodical question of the possibility of
formulation of the criteria of the non-existence, existence and/or
uniqueness of a shared metric $\Theta$, for some two pre-selected
candidates for physical quantum observables $A$ and $B$ at least.

\section{Merits of perturbative considerations\label{pet}}

Using a toy model we re-confirmed, in subsection \ref{sec1a}, the
well known fact that even in the case of the single given observable
$A \neq A^\dagger$ with real spectrum and even in the not too
complicated models with an infinitesimally small non-Hermiticity the
construction of a correct Hermitizing physical metric $\Theta$ may
be a formidable task.

\subsection{Models with bounded-operator observables $A$, $B$,
\ldots\label{petar}}

The existence, number and construction of the solutions $\Theta$
specified by Eqs.~(\ref{nezarobotok}) or (\ref{16}) + (\ref{17})
will all vary with the dynamical input $A$, $B$, \ldots . One of the
best analyses of these possibilities was presented in \cite{Geyer},
i.e., paradoxically, in one of the oldest papers published on the
subject. The assumptions made in {\it loc. cit.} (and, in
particular, the restriction of the scope of the paper to the bounded
operators of observables) may be greeted (it renders the mathematics
entirely reliable and rigorous) as well as damned (because the
assumption excludes many models due to the unbounded nature of their
observables).

\subsection{$N$ by $N$ complex-matrix observables $A$, $B$, \ldots \label{petas}}

The analysis of the generic $A-B$ compatibility conditions
(\ref{nezarobotok}) or (\ref{16})+(\ref{17}) remains far from easy
even in the Hilbert spaces of a finite dimension $N<\infty$. This
analysis may be assisted by the computers. Various
finite-dimensional complex-matrix special cases of Eqs.~(\ref{16})
and (\ref{17}) may be considered as approximating realistic
scenarios. In the most common (and suitably truncated)
harmonic-oscillator basis, both of our respective toy-model
exemplifications (\ref{a5}) and (\ref{a}) of $A_0$ and $B_0$ will be
sparse matrices.

The truncation of bases could clarify the questions of the existence
of the metric in the limit $N \to \infty$. In the hypothetical case
of an affirmative answer, the recipe could render the construction
of $F$ feasible. The analysis could be also performed in an opposite
direction, i.e., the existence of the metric may be required in
advance. Then one can deduce the necessary conditions and
constraints imposed upon the pair of perturbations $A_1$ and $B_1$.

It makes sense to split the complex $N$ by $N$ matrices in the real
and imaginary parts. Once we use subscripts $_s$ and $_a$ marking,
respectively, the symmetric and antisymmetric real matrices, the
input information encoded in the Hermitian, complex  $N$ by $N$
matrices
 \be
 A_0=A_s+{\rm i}A_a\,,
 \ \ \
 B_0=B_s+{\rm i}B_a\,,
 \ \ \
 R=R_s+{\rm i}R_a\,,
 \ \ \
 S_0=S_s+{\rm i}S_a\,
 \ee
will generate the desirable ultimate Hermitian matrix solution
$F=F_s+{\rm i}F_a$ via Eqs.~(\ref{16})+(\ref{17}), i.e., with
commutators in the real-matrix relations
 \be
 [A_s,F_s]-[A_a,F_a]=-R_a\,
 \ \ \ \
 [A_a,F_s]+[A_s,F_a]=R_s\,
 \label{27}
 \ee
and
 \be
 [B_s,F_s]-[B_a,F_a]=-S_a\,
 \ \ \ \
 [B_a,F_s]+[B_s,F_a]=S_s\,.
 \label{28}
 \ee
We may recall the symmetries/antisymmetries of matrices and omit the
diagonal (i.e., trivially satisfied) part of the first items in both
Eqs.~(\ref{27}) and (\ref{28}). Using an arbitrary ordering of all
of the independent and nontrivial matrix elements this enables us to
re-arrange the upper triangular part of all of the $_a-$subscripted
upper-case real and antisymmetric $N$ by $N$ matrices into the
respective $M-$dimensional lower-case column vectors with
$M=N(N-1)/2$ (i.e., we replace $R_a$ by, say, $\vec{r}^{(M)}$, etc).
Similarly,  with $V=N(N+1)/2$ we compress the information carried by
the real and symmetric  upper-case $N$ by $N$ matrix $R_s$ to a
lower-case vector $\vec{r}^{(V)}$. We take the upper triangular part
of any $_s-$subscripted matrix and we replace it by its real
$V-$dimensional column-vector representation.

The procedure eliminates the redundancy and preserves the linearity
of Eqs.~(\ref{27}) and (\ref{28}). Using the self-explanatory
abbreviation for commutators we may finally convert the equations
into their respective compact final versions
 \be
  L^{(VV)}_{(A)}\vec{f}^{(V)}
 +L^{(VM)}_{(A)}\vec{f}^{(M)}
 =\vec{r}^{(V)}\,,
 \ \ \ \
 L^{(MV)}_{(A)}\vec{f}^{(V)}
 -L^{(MM)}_{(A)}\vec{f}^{(M)}
 =-\vec{r}^{(M)}\,
 \label{27lc}
 \ee
and
 \be
  L^{(VV)}_{(B)}\vec{f}^{(V)}
 +L^{(VM)}_{(B)}\vec{f}^{(M)}
 =\vec{s}^{(V)}\,,
 \ \ \ \
 L^{(MV)}_{(B)}\vec{f}^{(V)}
 -L^{(MM)}_{(B)}\vec{f}^{(M)}
 =-\vec{s}^{(M)}\,,
 \label{28lc}
 \ee
i.e., in the partitioned block-matrix notation,
 \be
 \left (
 \begin{array}{cc}
 L^{(VV)}_{(A)}&L^{(VM)}_{(A)}\\
 -L^{(MV)}_{(A)}&L^{(MM)}_{(A)}
 \ea
 \right )
 \left (
 \begin{array}{c}
 \vec{f}^{(V)}\\
 \vec{f}^{(M)}
 \ea
 \right )=
\left (
 \begin{array}{c}
 \vec{r}^{(V)}\\
 \vec{r}^{(M)}
 \ea
 \right )
 \label{31}
 \ee
and
 \be
 \left (
 \begin{array}{cc}
 L^{(VV)}_{(B)}&L^{(VM)}_{(B)}\\
 -L^{(MV)}_{(B)}&L^{(MM)}_{(B)}
 \ea
 \right )
 \left (
 \begin{array}{c}
 \vec{f}^{(V)}\\
 \vec{f}^{(M)}
 \ea
 \right )=
\left (
 \begin{array}{c}
 \vec{s}^{(V)}\\
 \vec{s}^{(M)}
 \ea
 \right )\,.
 \label{32}
 \ee
Equations (\ref{16}) and (\ref{17}) {\it alias} (\ref{31}) and
(\ref{32}) may be finally re-read as a pair (or as a multiplet) of
the real and linear matrix relations
 \be
 {\cal A} \vec{f}= {\rm i}\,\vec{r}\,,\ \ \ \ \ \
 \widetilde{\cal A} \vec{f}= {\rm i}\,\widetilde{\vec{r}}\,,
 \ \ \ \ \ \ \ldots
 \label{equus}
 \ee
containing the same vector $ \vec{f}$. By construction, the $N^2$ by
$N^2$ matrices ${\cal A}$ and $\widetilde{\cal A}$ (etc) are all
real.

Both the existence and non-existence of the real solution vector
$\vec{f}$ (with $N^2$ components) remains admitted by the
first-order perturbation approach. No qualitative change in the
conclusions is detected when one simplifies the mathematics and when
one moves from the general case to the scenario with infinitesimally
small non-Hermiticites. The transition from existence to
non-existence of the metric depends on the dynamical input encoded
into matrices ${\cal A}$ and ${\cal B}$ as well as into vectors
${\vec{r}}$ and ${\vec{s}}$.

The former, large-matrix part of the encoded dynamical input is
determined by the self-adjoint zero-order components of the
observables in Eqs.~(\ref{21}) and (\ref{22}). The second, vectorial
part of the input carrying the information about the
non-Hermiticites seems more compact.  This feature of
Eqs.~(\ref{equus}) should be attributed to the neglect of the
higher-order ${\cal O}(\varepsilon^2)$ terms in Eqs.~(\ref{21}) and
(\ref{22}).

\subsection{Illustration: $N=2$\label{petant}}

Let us employ the row-wise vectorial compactification of matrices at
the first nontrivial dimension $N=2$ with $V=3$,  $M=1$ and with the
replacements
 \be
 F_s
 =
 \left (
 \begin{array}{cc}
 x&z\\
 z&y
 \ea
 \right )\
 \to
 \vec{f}^{(3)}=\left(
 \ba
 x\\z\\y
 \ea
 \right )\,,
 \ \ \ \
 F_a
 =
 \left (
 \begin{array}{cc}
 0&p\\
 -p&0
 \ea
 \right )\
 \to
 \vec{f}^{(1)}=\left(
 \ba
 p
 \ea
 \right )\,
 \ee
etc. Then, the most general choice of the Hermitian part of the
dynamical input
 \be
  A_s
 =
 \left (
 \begin{array}{cc}
 a&c\\
 c&b
 \ea
 \right )\,,
 \ \ \ \ \
 A_a
 =
 \left (
 \begin{array}{cc}
 0&d\\
 -d&0
 \ea
 \right )\
 \ee
leads to the following explicit $N^2$ by $N^2$ matrix form
of Eq.~(\ref{27lc}),
 \be
 \left (
 \begin{array}{cccc}
 0&2d&0&-2c\\
 -d&0&d&a-b\\
 0&-2d&0&2c\\
 -c&a-b&c&0
 \ea
 \right )\,
 \left (
 \ba
 x\\z\\y\\p
 \ea
 \right )=
 \left (
 \ba
 r^{(3)}_1\\
 r^{(3)}_2\\
 r^{(3)}_3\\
 r^{(1)}_1
 \ea
 \right ).
 \label{jedenop}
 \ee
Once we compare the first and third line we may conclude that there
exist no nontrivial components $z$ and $p$ (i.e., the metric remains
trivial, diagonal) unless the non-Hermiticity $A_1$ in ansatz
(\ref{21}) and in relation (\ref{16}) satisfies the non-diagonality
constraint
 \be
 r^{(3)}_1\ (=(R_s)_{11}) =-r^{(3)}_3\ (=-(R_s)_{22}).
 \label{ndc}
 \ee
Under the latter assumption we have just
three linearly independent equations for
the four real unknowns $x,z,y$ and $p\,$ so that,
in accord with the expectations \cite{uniq},
the admissible two-by-two metrics will form a one-parametric family.

An analogous elementary analysis has to be applied in the situation
with $j_{max}=1$ in which, under the non-triviality assumption
(\ref{ndc}) (plus under its $B-$related analogue), the three
independent lines of Eq.~(\ref{jedenop}) become complemented by
their three independent $B-$related analogues marked, for the sake
of simplicity, by tildas. With the latter additional dynamical
information at our disposal we arrive at the set of six equations
 \be
 \left (
 \begin{array}{cccc}
 0&2d&0&-2c\\
 -d&0&d&a-b\\
 -c&a-b&c&0\\
 0&2\tilde{d}&0&-2\tilde{c}\\
 -\tilde{d}&0&\tilde{d}&\tilde{a}-\tilde{b}\\
 -\tilde{c}&\tilde{a}-\tilde{b}&\tilde{c}&0
 \ea
 \right )\,
 \left (
 \ba
 x\\z\\y\\p
 \ea
 \right )=
 \left (
 \ba
 r^{(3)}_1\\
 r^{(3)}_2\\
 r^{(1)}_1\\
 s^{(3)}_1\\
 s^{(3)}_2\\
 s^{(1)}_1
 \ea
 \right ).
 \label{pairofops}
 \ee
The values of $z$ and $p$ get evaluated most easily. After their
re-insertion (reflected by our adding a hat to the right-hand side
vector elements) we are left with the following four linear
equations for the last two unknowns,
 \be
 \left (
 \begin{array}{cccc}
 %0&2d&0&-2c\\
 -d&d\\
 -c&c\\
 %0&2\tilde{d}&0&-2\tilde{c}\\
 -\tilde{d}&\tilde{d}\\
 -\tilde{c}&\tilde{c}
 \ea
 \right )\,
 \left (
 \ba
 x\\y
 \ea
 \right )=
 \left (
 \ba
 \hat{r}^{(3)}_2\\
 \hat{r}^{(1)}_1\\
 \hat{s}^{(3)}_2\\
 \hat{s}^{(1)}_1
 \ea
 \right ).
 \label{rofops}
 \ee
We have to avoid the non-existence of the metric, i.e., we have to
impose the triple restriction
 \be
 \hat{r}^{(3)}_2/d=
 \hat{r}^{(1)}_1/c=
 \hat{s}^{(3)}_2/\tilde{d}=
 \hat{s}^{(1)}_1/\tilde{c}\,
 \ee
upon the dynamical input information. The nontrivial solvability is
then guaranteed while the one-parametric ambiguity of the shared
physical metric will survive. Its suppression would require either
the choice of $j_{max}>1$ or an inclusion of the second-order
perturbation corrections in $\varepsilon$.

\section{Summary\label{summary}}

After one requires that a given pair of operators $A$ and $B$ with
real and non-degenerate spectra represents two quantum observables,
the specification of the $S-$superscripted physical Hilbert space
via the definition of the metric $\Theta$ may be impossible, unique
or ambiguous. An optimal scenario will be only realized in the case
of uniqueness of the metric. Still, even in the presence of an
ambiguity the authors of review \cite{Geyer} argued that our
knowledge of {\em any} formally correct metric will be welcome,
e.g., for variational-calculation purposes.

Such a pragmatic approach was accepted in virtually all of the
related literature. Some of the authors insisted on the optimality
(i.e., uniqueness) of the metric \cite{uniq}. Often, they believed
that such a goal is rather easy to achieve. In our present paper we
demonstrated that it is not always so.

The reasons and consequences have been explained in detail. In
particular, we came to the conclusion that the straightforward, best
known and successful suppression of the ambiguity of the metric via
the identification of the second observable with a ``charge''
($B={\cal C}$ such that ${\cal C}^2=1$, cf. \cite{Carl} for all
details) was exceptional, having been only rendered possible due to
a number of additional, {\it ad hoc} assumptions.

One can characterize the popular choice of the charge ${\cal C}$ as
a special implementation of the general recipe given in
Ref.~\cite{Geyer} and requiring the irreducibility of the set $A$,
$B$, \ldots of the quasi-Hermitian representations of the physical
quantum observables. Using a number of toy models we explained why
the consistent coexistence of more than one preselected
non-Hermitian candidate for the observable $\Lambda_j$ should be
considered exceptional.

In other words, whenever one tries to work with the two or more
independent and manifestly non-Hermitian candidates for quantum
observables, the absence of the reliable (and, in general,
difficult!) proof of the existence of the shared metric $\Theta$
should not be tolerated because it might really very easily result
in the loss of the applicability of the entire sophisticated PTSQM
formalism.

\section*{Acknowledgement}

The participation of MZ was supported by IRP RVO61389005 and by the
GA\v{C}R grant Nr. 16-22945S. The participation of IS and FR was
supported by IRP RVO61389005 and by the CTU grants Nr.
SGS15/215/OHK4/3T/14 and SGS16/239/OHK4/3T/14. HM and IL appreciate
the financial support by the Algerian Ministry of education and research.

%\newpage

\end{document}